\documentclass[conference,a4paper]{IEEEtran}
\usepackage{xcolor}
\usepackage{balance}
\usepackage{xfrac}

\usepackage{float}
\usepackage{amsmath}
\ifCLASSOPTIONcompsoc
 \usepackage[caption=false,font=normalsize,labelfont=sf,textfont=sf]{subfig}
\else
 \usepackage[caption=false,font=footnotesize]{subfig}
\fi
\usepackage{url}
\usepackage{amssymb}

\begin{document}
%
\title{Design of a Dichroic Transmissive Huygens' Metasurface Unit-Cell Presenting Refraction Angle Duality}

\author{\IEEEauthorblockN{
Georgios Kyriakou\IEEEauthorrefmark{1}, 
Giampaolo Pisano\IEEEauthorrefmark{1}, 
Luca Olmi\IEEEauthorrefmark{2}, 
Francesco Piacentini\IEEEauthorrefmark{1}
}                                     
\IEEEauthorblockA{\IEEEauthorrefmark{1}
University of Rome `La Sapienza' Physics Department, Rome, Italy\\
\IEEEauthorrefmark{2}
INAF-Arcetri Astrophysical Observatory, Florence, Italy\\georgios.kyriakou@uniroma1.it, giampaolo.pisano@uniroma1.it, luca.olmi@inaf.it, francesco.piacentini@roma1.infn.it}
}



\maketitle

\begin{abstract}
A purely transmissive Huygens' metasurface model under plane-wave illumination is used to derive circuit parameters describing a constituent unit cell, such that diverse refraction angles are attained at two distinct frequency bands. Various levels of accuracy of the circuit description approaching the analytical are possible by constraining certain numbers of parameters. This theoretical study is then tested by calculating the exact formulas of the two representations for the various strategies proposed. By using simulations of a candidate unit-cell, we then examine whether such circuit parameters correspond to rudimentary versions of the geometry of a so-called parallel `dogbone' structure. A device of this type is intended as dual-band (dichroic), dual-angle beam refractor diverting an incoming beam at different directions in two different bands without reflections.
\end{abstract}

\vskip0.5\baselineskip

%

\section{Introduction}

Metasurface technology has in the last two decades advanced from early theoretical concepts to many practical applications \cite{maci2024}. Deeper theoretical understanding of the properties of such two-dimensional models has also progressed, both at the macroscopic \cite{epstein2016} and the microscopic level \cite{ghaneizadeh2024}. In particular, Huygens' metasurfaces use homogenised impedance boundary conditions \cite{ataloglou2021} to describe the reflection and transmission, which after field averaging can also lead to an equivalent collocated impedance-admittance sheet \cite{pfeiffer2013}, as well as a microwave network description in terms of scattering matrices \cite{epstein2016}. These formulas being generic under any phase profile of the incoming wave, one can impose any desired spatial and spectral wavefront variations to achieve certain phase control of the outcoming wave.

In many applications, it is desirable to have a device that can work in transmission, with minimal reflections, in two separate frequency bands, e.g. such a device has been presented in \cite{hecht2024}. In designing separate unit-cell lattices for each frequency band though, conditions relating to the isolation between two separate sheets must be met, which depend on higher-order Floquet-mode excitations. If a single unit cell instead has naturally dual-band behaviour approaching the ideal scattering matrix representation of the metasurface cell in these bands, the fine-tuning of one geometric configuration would be a simpler design procedure. Dual-band meta-beamsteering by means of a feed-illuminated device has also been reported in \cite{ahmed2024}, but that solution regards a mechanically rotable system.

In our present work, we aim not only to satisfy the transmission-only condition that has been a well-described goal of metasurface design \cite{asadchy2016}, but to refract the beam in diverse angles in two different target bands. We call such a device a {\it dichroic dual-angle refractor}. Building upon previous experience in the implementation of astronomical receiver-chain components based on the metal-mesh technology \cite{pisano2015}, such a device would be useful in dual-band radio astronomical receivers, when an incoming beam has to be diverted in separate back-end systems. Added advantages of such a component would be its size and modularity, avoiding multiple reflective/refractive surfaces which necessitate trickier integration, as well as the attenuation of spurious effects such as Wood's anomalies \cite{falco2004}.

\section{Unit cell Impedance matrix structure and Second-order LC circuit approximation \label{sec:theory}}

In works of the last decade \cite{epstein2014,asadchy2016}, the macroscopic characteristics of a planar sheet to achieve reflectionless transmission and control of the outcoming wavefront phase have been derived as a combination of continuous impedance and admittance that it should present to an illuminating wave inducing electric and magnetic fields, respectively. This model also has a circuit equivalent, as outlined in \cite{selvanaygam2013}. 
\begin{figure}[h]
    \centering
    \includegraphics[width=\columnwidth]{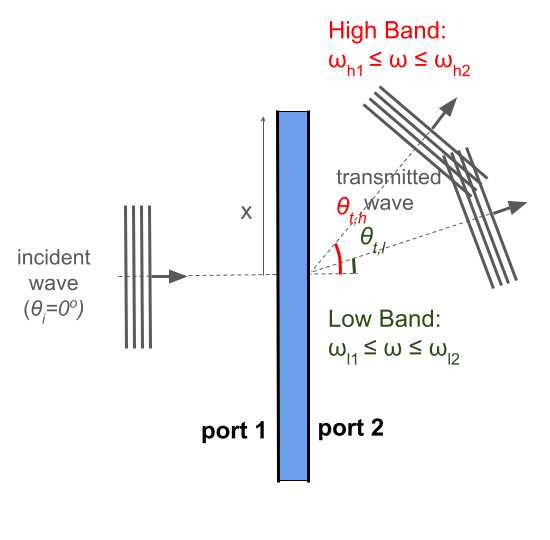}
    \caption{Schematic of the operating principle of a dual-band metasurface achieving transmission angle duality in two different frequency bands.}
    \label{fig:schematic}
\end{figure}

For such an ideal, continuous and reciprocal Huygen's metasurface, it is proven that when the incident angle of the {transverse electric} power wave is impinging at \( \theta_i=0^\circ \), the impedance matrix is given by \cite{epstein2014}:
\begin{align}
    Z_{11}(\omega,x)=Z_{22}(\omega,x)&=-j(\eta_0/\cos{\theta_t})\cot{(\phi(\omega,x))} \label{eq:Z11} \\
    Z_{12}(\omega,x)=Z_{21}(\omega,x)&=-j(\eta_0/\cos{\theta_t})\csc{(\phi(\omega,x))} \label{eq:Z12}
\end{align}
where \( \phi(\omega,x)=\omega x \sin{\theta_t}/c_0 \), \( \eta_0,\ c_0 \) the free space impedance and velocity of light respectively, while \( \omega \) is the angular frequency and \( x \) is the distance of the cell from the phase center. With this definition, the phase \( \phi(\omega,x) \) of the trigonometric functions of Eqs.~(\ref{eq:Z11}),~(\ref{eq:Z12}) corresponds to a planar wavefront dependence of a monochromatic (non-dispersive) wave which is refracted at angle \( \theta_t \). Here we use the simpler definition that does not take into account the different wave impedances of incident and refracted wave in order to preserve the symmetry property; study of the asymmetric impedance matrix as defined by \cite{wong2016,asadchy2016} will be conducted in future work.

Our goal is to achieve a duality of refraction angles at two separate frequency bands, a low-band (LB) \( \omega\in[\omega_{{l_1}},\ \omega_{{l_2}}] \) and a high-band (HB) \( \omega\in[\omega_{{h_1}},\ \omega_{{h_2}}] \). Let us denote \( \theta_{t,l},\ \theta_{t,h}\in [-90^\circ,\ 90^\circ] \) these two angles. Fig.~\ref{fig:schematic} shows a schematic of the operating principle of such a Huygen's metasurface. The ports are chosen as planes perpendicular to the incident wave propagation direction, and as long as they enclose the geometry, any free-space distance can be de-embedded from the network parameters. The transmission angles have here been drawn on the same quarter-plane since this is the solution obtained, as will later be shown.


The cotangent and cosecant functions have the following Laurent series representations around their common singular points \( \omega_s=s\pi c_0/(x\sin{\theta_t}),\ s=s(x) \in\{\hdots,-1,0,1,\hdots\} \):
\begin{align}
   \cot{\left(\phi(\omega,x)\right)} & \approx \frac{1}{\phi(\omega-\omega_s,x)}-\frac{\phi(\omega-\omega_s,x)}{3} \label{eq:cot} \\
   \csc{\left(\phi(\omega,x)\right)} & \approx \frac{1}{\phi(\omega-\omega_s,x)}+\frac{\phi(\omega-\omega_s,x)}{6} \label{eq:csc}
\end{align}
when \( |\omega-\omega_s| \) is small or practically when \( \phi(\omega-\omega_s,x)\lesssim \pi/4 \). This truncation allows for a relative error of less than \( 10^{-2}\) for all such angles. We could therefore reasonably replicate this behaviour (a rational function of maximum second order) instead of the analytic function, by employing a circuit model of second order. 

Let us now consider a collocation of two non-lossy resonant LC circuits, one in series and one in parallel, which represent an electric (\( \omega_e \)) and magnetic (\( \omega_m \)) resonance of a unit cell respectively. Such a unit cell has been proposed in \cite{capolino2013}, as we will later in more depth investigate. According to the X-equivalent circuit of \cite{selvanaygam2013}, the self-impedance of the unit cell is approximated as the mean value of the two LC circuit impedances. If we omit for brevity the functional dependence on the distance \( x \), we have:
\begin{equation} 
    Z_{11}^{\rm uc}(\omega)=\frac{j}{2}\left(\frac{\omega L_m}{1-(\omega/\omega_m)^2}-\frac{1-(\omega/\omega_e)^2}{\omega C_e} \right) 
    \label{eq:Zuc11}
\end{equation}

At the low-frequency region \( \omega \ll \omega_e,\omega_m \) we can derive the approximation:
\begin{equation} 
    Z_{11}^{\rm uc}(\omega)\approx \frac{j}{2}\left(\omega L_m-\frac{1}{\omega C_e} \right) 
    \label{eq:Zuc11_lowfreq}
\end{equation}

If we map this region to the one representing the right-most limit of the singular point \( \omega_0=0 \) of the cotangent, that is \( \omega\rightarrow 0^+ \), then using Eqs.~(\ref{eq:Z11}),~(\ref{eq:cot}),~(\ref{eq:Zuc11_lowfreq}), we can equate the coefficients of the \( 1/\omega,\ \omega \) terms imposing the \( \theta_{t,l} \) angle to get:
\begin{align}
    C_e=&\frac{\sin{(2\theta_{t,l})}x}{4\eta_0 c_0}
    \label{eq:Ce} \\
    L_m=&\frac{2\eta_0x\tan{\theta_{t,l}}}{3c_0} \label{eq:Lm}
\end{align}

{If \( x<0 \), the same phase shifts \( \phi \) modulo \( 2\pi \) are attainable such that blocks resembling the \( x>0 \) cells can be replicated without resulting in negative lumped element values}.

At the high-frequency region such that \( \omega+\omega_e\gg 1,\ \omega+\omega_m\gg 1 \) (but where \( |\omega-\omega_e|,\ |\omega-\omega_m| \) are still in the vicinity of 1, assuming that \( \omega_e,\ \omega_m \) are close to each other) we can derive the approximation:
\begin{equation} 
    Z_{11}^{\rm uc}(\omega)\approx \frac{j}{2}\left(\frac{1}{2C_m(\omega_m-\omega)} - 2L_e(\omega_e-\omega) \right) 
    \label{eq:Zuc11_highfreq}
\end{equation}

We now make the assumption that the singular point of order \( s(x) \) for the unit-cell at position \( x \) is such that \( \omega_{s(x)}\in [\omega_{h_1}\  \omega_{h_2}] \). We are therefore interested in mapping the response around \( \omega-\omega_{{s(x)}}\rightarrow 0^- \) to the Laurent approximation of Eq.~(\ref{eq:cot}). To do this we equate the inverse frequency coefficients of Eqs.~(\ref{eq:Zuc11}),~(\ref{eq:Zuc11_highfreq}) as well as the singular points of the two expressions, imposing a \( \theta_{t,h} \) angle, to get, after some algebra:
\begin{align}
    \omega_m &= \frac{s(x)\pi c_0}{\sin{\theta_{t,h}}x} \label{eq:omega_m} \\
    C_m &= \frac{\sin{(2\theta_{t,h})}x}{8c_0\eta_0}
    \label{eq:Cm}
\end{align}

{From Eq.~(\ref{eq:omega_m}), since we would like to keep the resonances close to each other for all \( x \) within a certain radius, a choice of \( s(x)\propto\lceil x \rceil \) should be made.} Eq.~(\ref{eq:Cm}), when compared to Eq.~(\ref{eq:Ce}), leads to \( C_e\neq 2C_m \) such that the refraction angles be \( \theta_{t,l}\neq \theta_{t,h} \) in the low and high frequency bands, respectively. We also match the coefficients and pivot points of the linear terms of Eqs.~(\ref{eq:Zuc11}),~(\ref{eq:Zuc11_highfreq}) to get:
\begin{align}
    \omega_e &= \frac{s(x)\pi c_0}{x\sin{\theta_{t,h}}}=\omega_m \label{eq:omega_e} \\
    L_e &= \frac{\eta_0x\tan{\theta_{t,h}}}{3c_0}
    \label{eq:Le}
\end{align}

Again, from Eq.~(\ref{eq:Lm}),~(\ref{eq:Le}), \( L_m\neq 2L_e\) in order to have \( \theta_{t,l}\neq\theta_{t,h} \). From the above Eqs.~(\ref{eq:Ce}),~(\ref{eq:Lm}),~(\ref{eq:Cm}),~(\ref{eq:Le}) all four lumped element parameters of the equivalent network are calculated by means of \( \theta_{t,l},\ \theta_{t,h} \). Eqs.~(\ref{eq:omega_m}),~(\ref{eq:omega_e}) first imply that the resonant frequencies are equal, as well as that the two angles are not free to choose, since the resonant frequencies are given by \( \omega_m=(L_mC_m)^{-1/2},\ \omega_e=(L_eC_e)^{-1/2} \). These equations, if substituted by Eqs.~(\ref{eq:Lm}),~(\ref{eq:omega_m})~(\ref{eq:Cm}) or Eqs.~(\ref{eq:Ce}),~(\ref{eq:omega_e})~(\ref{eq:Le}) respectively and worked out algebraically, lead to the conditions:
\begin{align}
    \frac{\tan{\theta_{t,h}}}{\tan{\theta_{t,l}}}&=\frac{(s(x)\pi)^2}{6} \label{eq:omegam_condition} \\
    \frac{\sin{(2\theta_{t,h})}}{\sin{(2\theta_{t,l})}}&=\frac{(s(x)\pi)^2}{6}
    \label{eq:omegae_condition}
\end{align}

Clearly these two conditions cannot be both satisfied, and can only be both approximated with low error when \( \theta_{t,l},\ \theta_{t,h}\) are small angles. Then we would get the unique condition:
\begin{equation}
    \frac{\theta_{t,h}}{\theta_{t,l}}=\frac{(s(x)\pi)^2}{6} \label{eq:unique_condition}
\end{equation}

Another strategy to avoid the ambiguity arising from Eqs.~(\ref{eq:omegam_condition}),~(\ref{eq:omegae_condition}) is to place \( \omega_e \) away from \( \omega_m \) at a higher frequency band, such that the second term of Eq.~(\ref{eq:Zuc11_highfreq}) does not yet become a linear one, but remains approximated by \( -1/(C_e\omega) \) as per the second term of Eq.~(\ref{eq:Zuc11_lowfreq}). Then we only have the condition Eq.~(\ref{eq:omegam_condition}) to satisfy, and by arbitrarily choosing \( \omega_e \) we get \( L_e=1/(C_e\omega_e^2) \) discarding Eq.~(\ref{eq:Le}). Finally, since also the linear term \( L_m\omega \) is small at low frequencies, to avoid both constraints we could relax the choice of \( L_m \) by discarding Eq.~(\ref{eq:Lm}) and choosing \( L_m=1/(C_m\omega_m^2) \). Then the angles \( \theta_{t,l},\ \theta_{t,h} \) can arbitrarily be chosen. This will of course induce more error in the approximation, which we will shortly examine.

It should be noted that similar equations can result from matching the \( 1/\omega \) coefficients of \( Z_{12},\ Z_{12}^{\rm uc} \), where the \( Z_{12}^{\rm uc} \) is given as the halved differnece of the magnetic LC circuit impedance from the electric LC circuit impedance \cite{selvanaygam2013}: 
\begin{equation}
    Z_{12}^{\rm uc}(\omega)=\frac{j}{2}\left(\frac{\omega L_m}{1-(\omega/\omega_m)^2}+\frac{1-(\omega/\omega_e)^2}{\omega C_e} \right)
    \label{eq:Zuc12}
\end{equation}

Since the factor of the linear term of Eq.~(\ref{eq:csc}) is \( 1/6 \), Eqs.~(\ref{eq:Lm}),~(\ref{eq:Le}),~(\ref{eq:unique_condition}) would have to be multiplied by \( 1/2 \). Instead of having to choose between minimizing the error of the \( Z_{11} \) or \( Z_{12} \) approximation by using either the \( 1/3 \) or \( 1/6 \) factor respectively, we could choose the linear slope that bifurcates these as slopes of a line, that is \( \alpha=\tan{(1/2(\arctan{(1/3)}+\arctan{(1/6)})}=0.2484\approx 1/4 \).  

{Finally, we point out that the distribution of unit cells is determined by sampling the \( x\)-axis of the metasurface, having first determined the unit-cell size \( D_{\rm uc} \) and the metasurface length \( D_x \). This sampling is then expressed as \( x_n=D_{\rm uc}/2+nD_{\rm uc},\ n\in\{1,\hdots,\lfloor D_x/D_{\rm uc} \rfloor\}\). This will of course introduce errors in the full-metasurface performance, which are assessed as part of ongoing simulations.}

\begin{figure}
    \centering
    \includegraphics[width=\columnwidth]{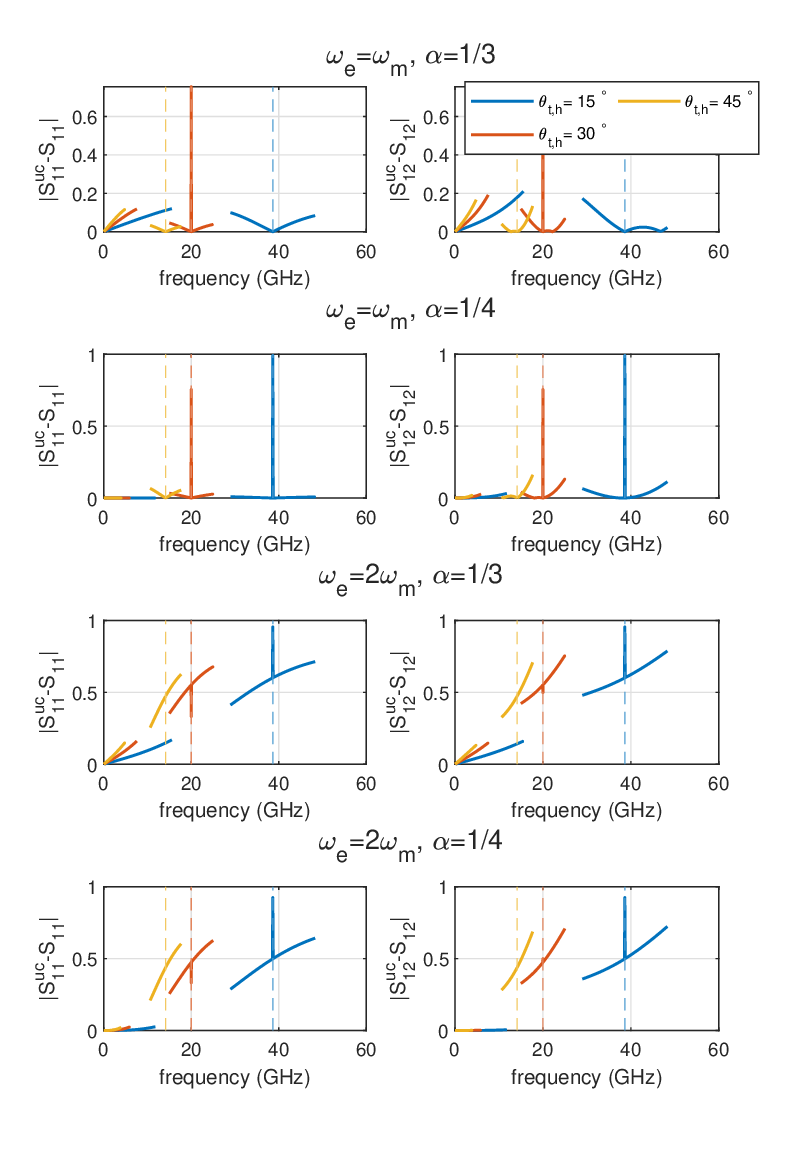}
    \caption{Relative error of the approximations \( S_{11}\approx S_{11}^{\rm uc}\) (left panels) and \( S_{12}\approx S_{12}^{\rm uc}\) (right panels). Various strategies are used, by means of the electric resonance position \( \omega_e=n\omega_m,\ n\in\{1,2\} \) and the factor \( \alpha \) used in Eqs.~(\ref{eq:Lm}),~(\ref{eq:Le}),~(\ref{eq:omegam_condition}).}
    \label{fig:rel_error_Zparams}
\end{figure}

To test the various methodologies presented above, we numerically calculate the two full representations using Eqs.~(\ref{eq:Z11}),~(\ref{eq:Z12}) for the analytic model and Eqs.~(\ref{eq:Zuc11}),~(\ref{eq:Zuc12}) for the circuit equivalent, transforming to S-parameters. We test three pairs of LB/HB angles, choosing \( \theta_{t,h}\in\{15^\circ,30^\circ,45^\circ\} \) and calculating the corresponding \( \theta_{t,l}\) from Eq.~(\ref{eq:omegam_condition}). Also, a value of \( x=15\)~mm~\( =\lambda \) at 20~GHz is used for this example. For each of the cases \( \omega_e=\omega_m \) and \( \omega_e=2\omega_m\), we employ the factors \( a=1/3 \) and \( a=1/4 \) in Eqs.~(\ref{eq:Lm}),~(\ref{eq:Le}),~(\ref{eq:omegam_condition}) (Eq.~(\ref{eq:Le}) is used only when \( \omega_e=\omega_m \), as per the previous). Fig.~(\ref{fig:rel_error_Zparams}) shows the absolute errors \( |S_{11}^{\rm uc}-S_{11}|\) (left panels) and \( |S_{12}^{\rm uc}-S_{12}| \) (right panels) for all these cases. The bands are not strictly chosen with respect to an error threshold, but are indicatively limited to \( \phi(x,\omega_s) \pm\pi/4 \) of each singular point (vertical dashed lines). Some high values on or around the singular point are attributed to numerical difficulties in approximating fractions of the type \( \infty/\infty \). LBs and HBs in the figure are for each angle recognised by the same color.

It is seen that the strategy of shifting upwards \( \omega_e \) to \( 2\omega_m \) increases the error in the approximation, while the factor \( \alpha=1/4 \) indeed lowers \( |S_{12}^{\rm uc}-S_{12}| \) while apparently also \( |S_{11}^{\rm uc}-S_{11}| \). Interestingly, higher angles seem to lead to less error but do also reduce the bandwidth. The best set of parameters would have to be chosen on the basis of an error threshold.

\begin{figure}[H]
    \centering
    \includegraphics[width=\columnwidth]{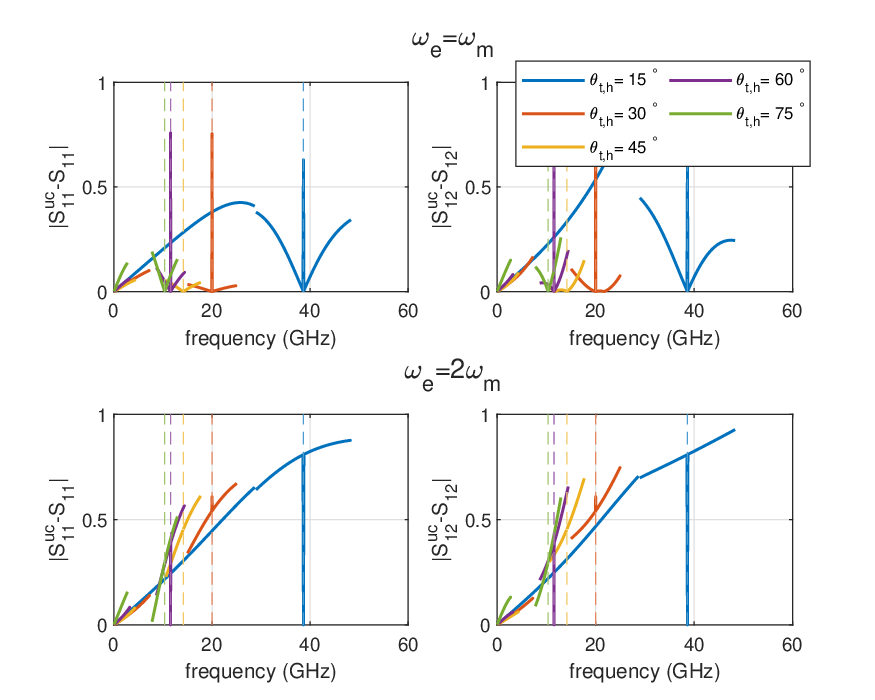}
    \caption{Relative error of the approximations \( S_{11}\approx S_{11}^{\rm uc}\) (left panels) and \( S_{12}\approx S_{12}^{\rm uc}\) (right panels) for \( \omega_e=n\omega_m,\ n\in\{1,2\} \). The constraint of Eq.~(\ref{eq:omegam_condition}) is not used here; it is instead chosen that \( \theta_{t,l}=\theta_{t,h}-10^\circ\).}
    \label{fig:rel_error_unconstrained}
\end{figure}

In all of these cases, the maximum \( \Delta\theta_t=\theta_{t,h}-\theta_{t,l} \) is around \( 6^\circ \). Such a slight difference presents practical challenges for the application design of such a Huygens' metasurface. We therefore also tested the same two cases of \( \omega_e=\omega_m \) and \( \omega_e=2\omega_m \) where now Eqs.~(\ref{eq:Lm}),~(\ref{eq:Le}),~(\ref{eq:omegam_condition}) are all discarded, while for an extended set of angles \( \theta_{t,h}\in\{15^\circ,30^\circ,45^\circ,60^\circ,75^\circ\} \) we choose a minimum difference of \( 10^\circ \) such that \( \theta_{t,l}=\theta_{t,h}-10^\circ\). Fig.~\ref{fig:rel_error_unconstrained} shows the absolute S-parameter errors for this strategy.

In this figure, the errors are higher than previous cases. Even though it still seems that \( \omega_e=2\omega_m\) leads to higher errors, the case \( \omega_e=\omega_m \) also leads to erroneous approximation especially for \( \theta_{t,h}=15^\circ \). For this angle, the LB and HB almost overlap. For this strategy, it seems that the case \( \omega_e=\omega_m \) with an angle \( \theta_{t,h}=30^\circ \) can offer a compromise between low relative error (\( <0.2\)) and acceptable bandwidth. 

Given the difficulties in approximating a one-layer metasurface analytic expression, we aim to examine the same problem by introducing more layers separated by dielectric patches in the future.

\section{Preliminary Simulations and Circuit Parameter Fitting of Unit-Cell Geometry}
\begin{figure}[ht]
    \centering
    \includegraphics[width=\columnwidth]{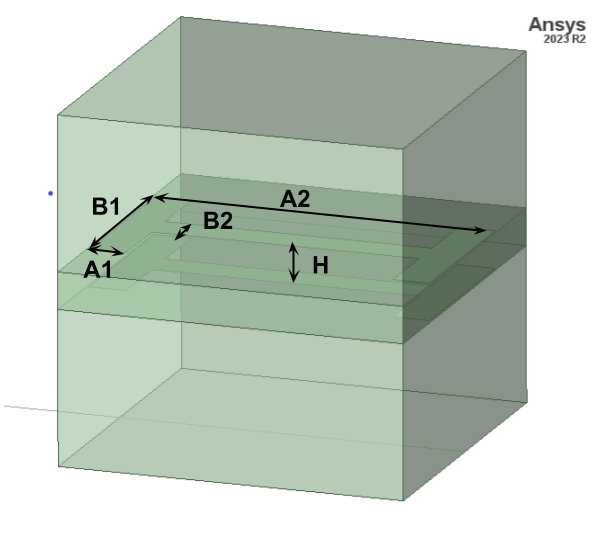}
    \caption{Parallel `dogbone' unit cell geometry depicted in Ansys HFSS. The various strip lengths are denoted, in accordance with \cite{capolino2013}. The darker region is filled with a dielectric material of \( \epsilon_r=3.48 \).}
    \label{fig:hfss_geometry}
\end{figure}

In this work, the parallel `dogbone' geometry is examined as a potential unit cell of the dichroic dual-angle refractor. A screen capture of this geometry with the HFSS modeller can be seen in Fig.~\ref{fig:hfss_geometry}. Keeping the notation from \cite{capolino2013}, we highlight the dimensions of this cell on the same figure. In \cite{donzelli2012,capolino2013}, it has been proven that an equivalent circuit is composed by just the two circuits we have examined (and used to define Eqs.~(\ref{eq:Zuc11}),~(\ref{eq:Zuc12})) in Sec.~\ref{sec:theory}. This approximation derives from the treatment of the stripes as transmission line segments, and their dimensions being subwavelength with respect to the operating frequency bands.  

Using Ansys HFSS, we simulated instances of this geometry by varying the parameters A2 from 7.4~mm to 7.5~mm with a 0.05~mm step, as well as B1 from 3.5~mm to 4.5~mm with a 0.05~mm step. As Donzelli~\emph{et.~al.} point out in \cite{donzelli2012}, both resonances are affected by tuning these two parameters. We use \texttt{fmincon} in MATLAB to extract the parameters \( C_e,\ L_m,\ f_e,\ f_m\) by fitting both Eqs.~(\ref{eq:Zuc11}),~(\ref{eq:Zuc12}) to their simulated frequency response, using the sum of absolute relative errors as a cost function\footnote{To improve the \texttt{fmincon} convergence, the relative error does not have a constant denominator across frequency points but the maximum of the differentiated terms, such that each summed fraction is always within [0 1].}. Each current solution is used as an initial guess for the next fitting of the unit cell with closest A2, B1 values to those currently solved.

\begin{figure}[H]
    \centering
    \includegraphics[width=\columnwidth]{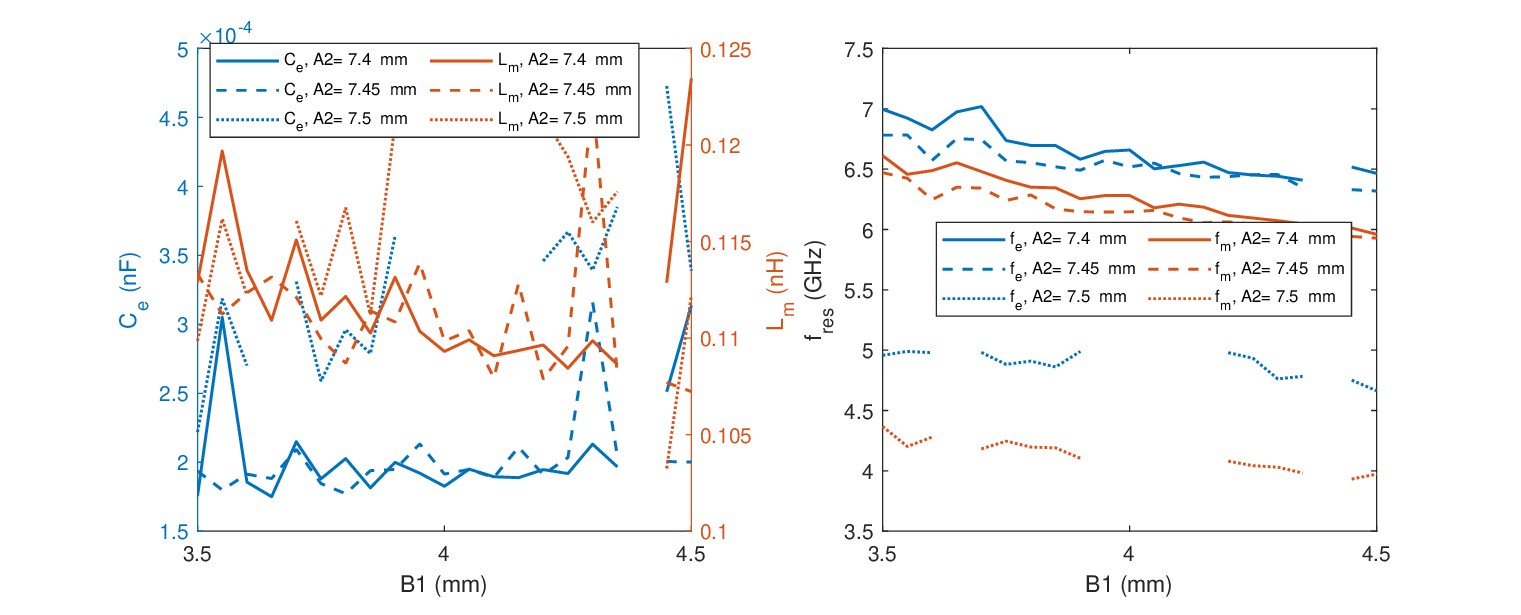}
    \caption{Left: \( C_e \) across B1 (left axis, blue) and \( L_m \) across B1 (right axis, orange) for 3 close values of A2, Right: \( f_e, \ f_m \) across B1 for the same A1 values. Omitted values presented high fitting error.}
    \label{fig:fit_params}
\end{figure}

In Fig.~\ref{fig:fit_params}, left, the capacitances \( C_e \) (in nF) and inductances \( L_m \) (in nH) across B1 for the 3 values of A2 are seen (with separate \( y\)-axes), while on the right the resonances \( f_e,\ f_m\) are plotted for the same parameter space. We demand a 10\% error threshold for the S-parameter fitting and omit the rest of the values. As can be seen, the parameters have a negative slope with respect to increasing B1 but \( C_e,\ L_m \) significantly deviate from a purely linear behaviour. The optimisation is sensitive to the initial guess, for which the gaps should be filled by other initialisation strategies or using global optimisation methods. Hints as to where the resonances lie are also given by the S-parameter local extrema.


{Finally, it should be noted that this parametric analysis is not exhaustive, since the unit cell size \( D_{\rm uc} \) as well as the separation between the two metallic structures \( H \) affect the resonant frequency magnitudes and their ratio, respectively. More extensive simulations are ongoing, and during the conference a better picture of the possibility of this unit cell, or adjustments of it, to cover the entire space of desirable phase shifts will be presented.
}


\section{Conclusions}
A dual-angle for dual-band transmissive Huygens' metasurface has been approximated by a collocation of two second-order electrically and magnetically resonant LC circuits. The approximation strategies are more successful when the two resonances coincide. A candidate unit cell satisfying the two-circuit collocation shows that this condition is not achieved by slight perturbations of a known geometry; future work will examine the veracity of this observation as well as the introduction of more metasurface layers to assist the attempted design.

\section{Acknowledgment}
We would like to thank Vasileios Ataloglou for his insightful comments that improved the quality as well as the scope of this paper. This work has been financially supported by Italian gonvernment funds from PRIN 2022 (Progetti di Ricerca di Rilevante Interesse Nazionale) grant no.~2022AYFS7F, as allocated through the postdoctoral research contract no.~0000924 of the Physics Department of `La Sapienza' University of Rome.

\balance

\end{document}